\documentclass[11pt]{article}
\usepackage{mathptm}
\usepackage[utf8]{inputenc}
\usepackage[T1]{fontenc}
\usepackage[english]{babel}
\usepackage{textcomp}
\usepackage{type1cm}
\usepackage[inline]{enumitem}
\usepackage{tabularx}
\newcolumntype{Y}{>{\centering\arraybackslash}X}

\let\oldbibliography\thebibliography
\renewcommand{\thebibliography}[1]{\oldbibliography{#1}
\setlength{\itemsep}{0pt}} 
    
\usepackage{hyperref}
\usepackage{graphicx}
\usepackage{listings}
\usepackage{subcaption}
\usepackage[acronym]{glossaries}
\newacronym{nilu}{NILU}{Norwegian Institute of Air Research}
\newacronym{met}{MET}{Norwegian Meteorological Institute}
\newacronym{pm}{PM}{Particulate Matter}
\newacronym{who}{WHO}{World Health Organization}
\newacronym{leo}{LEO}{Little Environmental Observatory}
\makeglossaries 

\begin{document}


\title{Teaching Electronics and Programming in Norwegian Schools Using the
\texttt{air:bit} Sensor Kit}

\author{
    Bjørn Fjukstad \\ Department of Computer Science \\ UiT The Arctic
    University of Norway  \and 
    Nina Angelvik\\ Department of Computer Science \\ UiT The Arctic
    University of Norway  \and 
    Morten Grønnesby\\ Department of Computer Science \\ UiT The Arctic
    University of Norway  \and 
    Maria Wulff Hauglann\\ Department of Computer Science \\ UiT The Arctic
    University of Norway  \and 
    Hedinn Gunhildrud \\ Science Centre of Northern Norway \and
    Fredrik Høisæther Rasch\\ Department of Computer Science \\ UiT The Arctic
    University of Norway  \and 
    Julianne Iversen \\ Faculty of Science and Technology \\ UiT The Arctic
    University of Norway \and
    Margaret Dalseng\\ Faculty of Science and Technology \\ UiT The Arctic
    University of Norway \and
    Lars Ailo Bongo \\ Department of Computer Science \\ UiT The Arctic
    University of Norway
}

\date{}
\maketitle

\section*{Abstract} 
We describe lessons learned from using the \texttt{air:bit} project to
introduce more than 150 students in the Norwegian upper secondary school to
computer programming, engineering and environmental sciences. In the
\texttt{air:bit} project, students build and code a portable air quality
sensor kits, and use their \texttt{air:bit} to collect data to investigate
patterns in air quality in their local environment. When the project ended
students had collected more than 400,000 measurements with their
\texttt{air:bit} kits, and could describe local patterns in air quality.
Students participate in all parts of the project, from soldering components
and programming the sensors, to analyzing the air quality measurements. We
conducted a survey after the project and describe our lessons learned from
the project. The results show that the project successfully taught the
students fundamental concepts in computer programming, electronics, and the
scientific method. In addition, all the participating teachers reported that
their students had showed good learning outcomes. 

\section*{Introduction}
We have developed the \texttt{air:bit}, an Arduino-based air quality sensor kit
that students build and program to collect air quality data in their local
environment \cite{airbit}. Together with the \texttt{air:bit} sensor kit, we
developed teaching materials that include how to assemble the \texttt{air:bit}
and how to program its different sensors, and a cloud based service for students
to upload and explore their collected datasets. All of which are openly
available online at \url{airbit.uit.no}. We used these resources develop an
interdisciplinary course for students in Norwegian upper secondary schools,
which we offered to different schools across Northern Norway. 

From our first deployment of the \texttt{air:bit} project we identified a set of
necessary improvements to scale out to even more schools. First, the
initial version of the sensor kit did not provide the students with comparable
data to official measurement stations. Second, because we wanted to invite more
schools we had to improve how we shared resources such as programming guides to
the students and teachers. Third, we had to improve the online platform for
exploring the datasets to handle data from many geographically distributed
locations.  Fourth, we needed to develop a survey to investigate the student
experiences to further improve the course and their learning outcomes.

In this paper we describe our experiences using the \texttt{air:bit} project in
11 Norwegian upper secondary schools. The participating schools were all in
Northern Norway, but spread over a large geographical area. We developed a
survey to collect information about the experiences and teaching outcome for
both students and teachers. We used this study to investigate the prior
knowledge about programming and electronics, the difficulties of the different
parts of the project, and to highlight areas of improvement. We describe how we
redesigned the sensor kit and its accompanying resources, the online data
exploration platform and the student experiences from this second deployment of
the \texttt{air:bit} project. 

We believe our work provides insight for others that provide similar
maker-inspired projects to students without any programming experience, and that
it answers some questions about the effect of such projects for motivating
students to engage in computer science and other STEM subjects.

\section*{The \texttt{air:bit} project}
We initiated the \texttt{air:bit} project in 2016 and have since offered it to
upper secondary schools at two separate occasions. First, in 2017 to a single
school, and then in 2018 to 16 schools from across Northern Norway. In our
previous publication, we describe the \texttt{air:bit} project, the first
version of our \texttt{air:bit} and our experiences from the first participating
class\cite{airbit}. Here we provide a short background in air pollution, details
of the second version of the \texttt{air:bit} sensor kit, and the other
necessary improvements needed to offer the course to more schools.

\subsection*{Motivation}
While programming or computational thinking is added to the school curricula
in countries such as the UK, Finland or Estonia, Norway is unfortunately falling
behind\cite{compfuture}. Initiatives such as Lær Kidsa
Koding\footnote{\url{kidsakoder.no}} are working with legislators to introduce
these concepts in the Norwegian educational system, but it is a time consuming
process. Fortunately, certain science subjects in the upper secondary school
allow teachers to add smaller projects that combine programming with the
traditional sciences. 

Air pollution is the largest single environmental health risk, and it
contributes to respiratory disease, cardiovascular disease and certain
cancers.\cite{who,guerreiro2013air,beelen2014effects,brook2010particulate,raaschou2013air,pascal2013assessing}
Air pollution originates from a wide range of sources, and especially
Particulate Matter (PM) from combustion engines and cars driving with studded
tires is a major problem in Norway\cite{guerreiro2013air,hagen2000associations}.

With cheaper and higher quality sensors, combined with easy-to-use
microcontrollers such the Arduino UNO, it is possible to develop small sensor
kits with little previous knowledge and experience with electronics.  Low-cost
air quality monitoring kits have already shown their effectiveness in citizen
science projects, by reducing the high cost of developing such measurement
stations.\cite{dutta2017towards, castell2015mobile, antonic2014urban,
oletic2015design}

\subsection*{\texttt{air:bit}} 
\begin{figure}[t]
    \centering
    \includegraphics[width=0.7\textwidth]{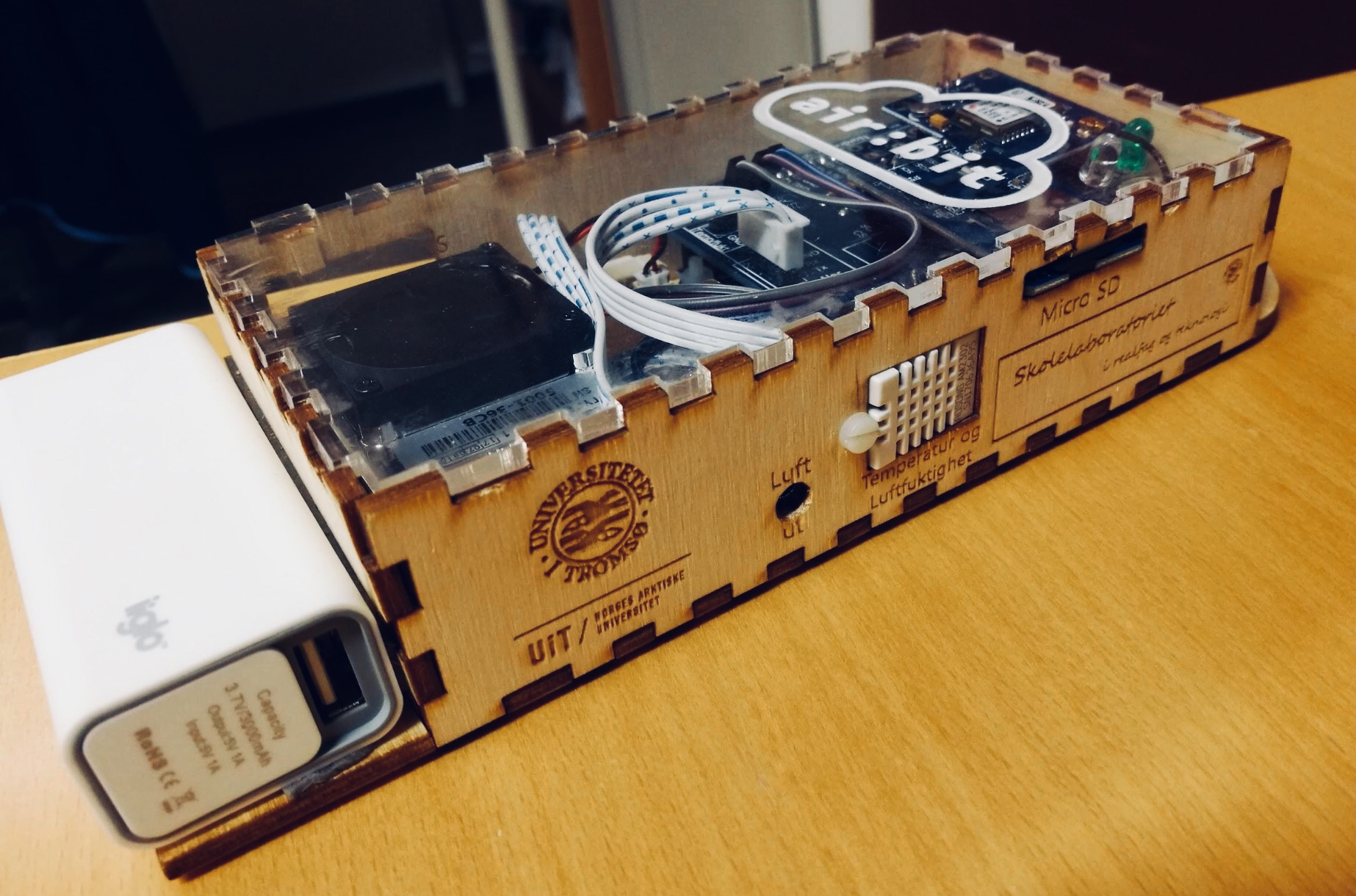} 
    \caption{The second version of the air:bit sensor kit.}
    \label{fig:kit} 
\end{figure}

The \texttt{air:bit} is a small microcontroller based data logger for measuring
dust particles, air temperature, air humidity, GPS-based location, time and
date. The kit is enclosed in a laser cut
box, equipped with an external battery for portability. Table
\ref{table:components} lists the different components and their respective cost,
and Figure \ref{fig:kit} shows the assembled kit.
We package and ship all components
to each school in cardboard boxes. 

We designed the kit as simple as possible to facilitate use in an educational
setting. To simplify the assembly and soldering of the components to the
microcontroller we use a custom PCB circuit board that has pre-defined pins for
each sensor, and fits on top of the Arduino UNO board. New in the second version
of our \texttt{air:bit} is the Nova SDS011 Dust sensor which allow students to
collect data that can be compared to official measurement stations (PM2.5 and
PM10). This sensor is more expensive than the previous dust sensor, but provides
higher quality data. We also update the temperature and humidity sensor from
the DHT11 to the DHT22. This sensor allows students to record temperatures below
0$^{\circ}$C, which is frequent in Northern Norway. We also re-designed the box
to make it easier for students to carry around. 

We continue to use the standard Arduino IDE together with additional libraries
to program the \texttt{air:bit}. The libraries provide the low-level
functionality to retrieve data from each sensor, and students use their
respective APIs to assemble a program that collects data from all sensors
simultaneously and write them to a memory card.  
We distribute example code to interface and collect data from the
individual sensors. These are small examples typically less than 100 lines of
code, and students typically end up with a complete solution of around 500 lines
of code. The data is recorded to the memory card using a simple CSV file format
that make it simple to view and inspect the output datasets.

\begin{table}[t]
    \centering
    \caption{A list of the different components in the air:bit
    along with their cost (as of January 2018).}
    \begin{tabular}{| l | l |}
        \hline 
        Component & Cost (USD) \\ \hline 
        Arduino Uno microcontroller & \$3.00 \\ 
        NEO6MV2 GPS module & \$6.00 \\ 
        Nova SDS011 dust sensor
        (PM2.5 and PM10).  & \$14.50 \\
        DHT22 temperature and humidity sensor & \$3.50 \\
        SD Card reader and 16GB memory card & \$9.00 \\
        Portable power bank & \$15.00 \\
        Custom PCB circuit board & \$3.00 \\
        Custom enclosure box & \$5.00 \\
        USB cable & \$2.00 \\
        Zip-lock bag with LEDs, resistors and spare parts & \$5.00 \\
        Cardboard box & \$1.50 \\
        \hline
        \textbf{Total:} & \$67.50 \\ \hline
    \end{tabular} 
    \label{table:components} 
\end{table} 

\subsection*{Improved Online Resources} 
To allow more schools to participate in the \texttt{air:bit} project we had to
improve the online resources.  We developed detailed guides on how to solder the
components of the \texttt{air:bit}, how to assemble the box, and how to program
each sensor.  We also included a resources page with video lectures on air
pollution from environmental researchers from the Norwegian Institute for Air
Research (NILU) and the The Norwegian Meteorological Institute (MET).  All of
these are available online at \url{airbit.uit.no}. These are open to everyone.

Together with the educational resources we developed the \texttt{air:bit}
platform, a system for storing, exploring and visualizing air quality data from
\texttt{air:bit} kits and other, external data sources.\cite{angelvik2018air,
angelvik2018data} The system consists of three individual parts, the
\texttt{air:bit} web application, a service for retrieving and uploading
datasets, and a backed storage service. 
The \texttt{air:bit} web application provides students with educational
resources such as instructions on how to solder each component and program the
sensors, in addition to an interactive visualizations of the air quality
datasets. The other two services are responsible for retrieving student data and
making them available through the interactive visualizations. We host the
backend services on Google Cloud Platform which makes it possible for us to
scale out if we get even more participating schools. Together with student data
we also retrieve data from NILU and MET through their open APIs. By developing
an online service that integrates the different data sources we make it easier
for students to explore the heterogeneous data.

\subsection*{The 2018 \texttt{air:bit} Project}
We have offered our \texttt{air:bit} course twice. The pilot was held in spring
2017 at the University of Tromsø to science students at a local High School
(Kongsbakken Vidergående Skole). The second round was in spring 2018 with 11
participating schools, and 164 students from across Northern Norway. In the
second round, the participating teachers chose to use the \texttt{air:bit}
project in four different subjects, from all three years of upper secondary
school. The 164 students built in total 62 \texttt{air:bit} sensor kits. 

We invited schools to participate in the project at the beginning of the fall
semester of 2017. The participating teachers were invited to a two-day workshop
where the goal was to teach them how to assemble and program the
\texttt{air:bit} sensor kits. 16 teachers participated and successfully built
and programmed 15 \texttt{air:bit} kits. Following the teacher workshops the
teachers themselves chose how to incorporate the \texttt{air:bit} project in
their classroom. 

In the spring of 2018 all participating classes were invited to attend a full
day workshop where they got help to finalize the assembly and programming of the
\texttt{air:bit} kits. Most classes only had minor programming issues, but some
student groups needed help to troubleshoot their soldering and assembly.
Following the workshop, each class returned home to start data collection. 

During the spring of 2018 students uploaded more than 400,000 unique
measurements to our online web application. The class with the largest number of
data points used their data to investigate and compare air pollution levels
on the outside playgrounds of local kindergartens. 

\section*{Evaluation}
We developed two surveys to investigate learning outcomes and experiences from
the \texttt{air:bit} course, one for students and another for teachers.
Specifically we wanted to investigate: 
\begin{enumerate*}[label=\roman*)] 
    \item the students prior experience with programming and electronics;
    \item how much time students spent working on the project;
    \item the overall difficulty of the project; and
    \item the programming specific learning outcomes, e.g. knowledge about
        variables and loops.
\end{enumerate*}

All questions were written in Norwegian, and we present their English
translation in this paper. We have also translated the free-text responses from
the students and teachers. We distributed both surveys online using a service
similar to Questback. 

\subsection*{Student Experiences} 
To investigate the student perceived learning outcome and experiences from the
course we asked the students to answer a questionnaire consisting of 14 multiple
choice questions, and 35 statements answered on a five-level Likert scale.  Of
the total 164 students, we received 90 individual responses. This gives 55\%
answer rate. Of the 90 responses 64\% were boys and 36\% girls. 

Table \ref{table:weeks} shows the reported number of weeks spent working on the
project and Table \ref{table:hours} shows the reported number of hours spent
every week working on the project. 

\begin{table}[!htbp] \centering 
    \caption{Reported number of weeks spent working on the \texttt{air:bit}
    project.} 
  \label{table:weeks} 
\begin{tabularx}{0.70\textwidth}{|Y|Y|Y|Y|Y|}
\hline
Less than 4 weeks & 4 to 8 weeks & 9 to 12 weeks & 13 to 16 weeks & More than 16
    weeks \\ 
\hline 
4\% & 20\% & 45\% & 27\% & 4\% \\ 
\hline 
\end{tabularx} 
\end{table} 

\begin{table}[!htbp] \centering 
  \caption{Reported hours spent working on the  \texttt{air:bit} project every week.} 
  \label{table:hours} 
    \begin{tabularx}{0.70\textwidth}{|Y|Y|Y|Y|}
    
\hline 
Less than 1 hour & 1 to 2 hours & 3 to 4 hours & More than 5 hours \\ 
\hline 
10\% & 23\% & 50\% & 17\% \\ 
\hline 
\end{tabularx} 
\end{table} 

We asked the students to evaluate the difficulty of several tasks in the
project. Table \ref{table:tasks} shows the questions and how the students
responded.  From the table we clearly see that the students found it easy to get
an overview of the components and solder the air:bit together. Further the
students fount the LEDs, the temperature and humidity sensor, and the dust
sensor relatively easy to program. However programming the GPS and memory card
was a more difficult task. The most difficult task of all was to assemble all
the individual parts into a single program. More than half the students answered
that it was either \emph{difficult} or \emph{very difficult}. 

\begin{table*}[!htbp] \centering 
  \caption{Survey results from reported difficulty in completing the different
    tasks of assembling and programming the \texttt{air:bit}.} 
  \label{table:tasks} 
        \begin{tabularx}{0.80\textwidth}{|Y|Y|Y|Y|} 
\hline 
\textbf{Statement} & \textbf{Very easy or easy} & \textbf{Nor easy nor
difficult} & \textbf{Difficult or very difficult} \\
\hline 
Get an overview of the components & 83\% & 13\% & 3\%\\
        \hline 
Solder the components & 80\% & 18\% & 2\% \\ 
\hline 
Program the LEDs & 73\% & 18\% & 9\% \\ 
\hline 
Program the temperature and humidity sensor & 50\% & 39\% & 11\% \\ 
\hline 
Program the dust sensor & 46\% & 41\% & 13\% \\ 
\hline 
Program the GPS & 20\% & 37\% & 43\% \\ 
\hline 
Program the memory card and memory card reader & 28\% & 42\% & 30\% \\ 
\hline 
Assemble the different programs into a single program & 15\% & 32\% & 53\% \\ 
\hline 
\end{tabularx} 
\end{table*} 

Further we asked the students to evaluate seven statements regarding their knowledge 
prior to starting the project.

\begin{table*}[!htbp] \centering 
  \caption{Survey results from the self-assessed general knowledge before and
    after participating in the \texttt{air:bit} project. We have combined the top and
    bottom categories.}
  \label{table:general-knowledge} 
        \begin{tabularx}{\textwidth}{|Y|Y|Y|Y|} 
\hline 
        \textbf{Statement} & \textbf{Very little or Little} & 
        \textbf{Neither little nor much} & \textbf{Much or Very much} \\
        \hline 
        How electrical circuits work &   
        \textit{32\%} $\rightarrow$ 17\% &
        \textit{39\%} $\rightarrow$ 33\% &
        \textit{29\%} $\rightarrow$ 50\%  \\ \hline
        How to solder electrical components &  
        \textit{23\%} $\rightarrow$ 8\%   &
        \textit{22\%} $\rightarrow$ 14\%  & 
        \textit{55\%} $\rightarrow$ 78\%  \\ \hline 
        How to write a computer program & 
        \textit{39\%} $\rightarrow$ 18\%    &
        \textit{38\%} $\rightarrow$ 32\% &
        \textit{23\%} $\rightarrow$ 50\%  \\ \hline
        What an Arduino is and how to program it & 
        \textit{46\%} $ \rightarrow$ 13\% 	&
        \textit{28\%} $ \rightarrow$ 28\% 	&
        \textit{26\%} $ \rightarrow$ 59\%  \\ \hline
        How to plan and execute a scientific project & 
       \textit{24\%} $\rightarrow$ 8\%   &
       \textit{29\%} $\rightarrow$ 30\%  &
       \textit{47\%} $\rightarrow$ 62\%  \\
 \hline 
        How to collect and analyze research data & 
        \textit{39\%} $\rightarrow$ 8\%   &
        \textit{31\%} $\rightarrow$ 30\%  & 
        \textit{30\%} $\rightarrow$ 62\%  \\
 \hline
        How to determine measurement uncertainty in research data &
        \textit{36\%} $\rightarrow$ 12\% & 
        \textit{41\%} $\rightarrow$ 38\% & 
        \textit{24\%} $\rightarrow$ 40\% \\
         \hline
\end{tabularx} 
\end{table*} 

\begin{table*}[!htbp] \centering 
\caption{
    Survey results from the self-assessed programming knowledge before and after
    participating in the \texttt{air:bit} project. We have combined the top and bottom
    categories.}
\label{table:coding-knowledge} 
\begin{tabularx}{\textwidth}{|Y|Y|Y|Y|} 
    \hline 
\textbf{Statement} & \textbf{Very little or Little} & \textbf{Neither
little nor much} & \textbf{Much or Very much} \\
\hline  
    What a variable is and how they are used &
  \textit{49\%} $\rightarrow$ 29\%   &
  \textit{17\%} $\rightarrow$ 28\%  & 
  \textit{34\%} $\rightarrow$ 43\%  \\ 
        \hline
    What a data type is and how they are used (e.g. \texttt{float} or
    \texttt{int}) & 
  \textit{50\%} $\rightarrow$ 31\%  & 
  \textit{26\%} $\rightarrow$ 28\%  & 
  \textit{24\%} $\rightarrow$ 41\%  \\
        \hline
    What a loop is and how they are used (e.g. a \texttt{for} loop) & 
  \textit{56\%} $\rightarrow$ 35\% & 
  \textit{22\%} $\rightarrow$ 24\% & 
  \textit{22\%} $\rightarrow$ 41\%\\
        \hline
    What a logic test is and how they are used (e.g. an \texttt{if} test) & 
  \textit{54\%} $\rightarrow$ 34\% & 
  \textit{20\%} $\rightarrow$ 25\% & 
  \textit{36\%} $\rightarrow$ 41\% \\
        \hline
    How to debug a computer program & 
    \textit{63\%}  $\rightarrow$ 30\% &  
   \textit{19\%}  $\rightarrow$ 29\% & 
   \textit{18\%}  $\rightarrow$ 41\% \\
\hline
\end{tabularx} 
\end{table*} 

94\% of the students reported that they used the online materials while they
where building and programming the \texttt{air:bit} . From these, 
83\% reported that these where either to a large degree or to a very large
degree helpful. 14\% reported that the online materials where to some extent
helpful, while 4\% reported that the material was not at all or to a small
degree helpful.

We asked the students to rate how satisfied or dissatisfied having participated
in the air:bit project. 69\% of the students reported that they were satisfied
or very satisfied, 23\% nor satisfied or dissatisfied and 8\% reported that they
were very dissatisfied or dissatisfied. 

We also left a text field where the students could enter free-text comments to
the project. In total 17 students used this opportunity to give us comments.
From these 17, two students wrote that they were given instructions
that the \texttt{air:bit} kits could be used outside, and that the kits had
malfunctioned from extensive outside exposure (5 days). Another student noted
that the \texttt{air:bit} box was a bit too small for all the components. 
One student reported "It is a difficult project, but it's fun". Another student
wrote "The project was very fun, but much of the work following the data
collection it was pretty gruesome to complete. However, I learned a lot of
lessons of how complicated such processes are". One student wrote that they did
not get the programming guidance they needed when they visited the university.
Another student wrote that the programming was too easy, and that it did not
challenge the students.  One student requested a guide on presenting research
data. 

\subsection*{Teacher Experiences} 
We developed a short survey for the teachers that where involved in the project.
Of the 11 participating teachers, 8 replied to our survey. From the responses we
got an overview of the time spent on the project, their knowledge about
programming and microcontrollers prior to the project, and the learning outcomes
for their students. The survey consisted of multiple choice questions and a
free text field where they could enter their own comments. 

Of the 8 replies, all teachers responded that their students had significant or
very significant learning outcomes. All teachers also responded that they would
recommend other classes to participate in the project. More than half of the
teachers reported that they had much, or very much knowledge about electrical
circuits, soldering, programming, and how to plan and execute a scientific
project. However, there were still teachers who reported that they had very
little or little experience in all of these areas. 

From the free text responses, one teacher reported that the students became a bit
tired towards the end due to the process of writing up their findings. One
teacher said that the motivation was highly variable between the students and
groups, and that they did not feel any ownership to the programming due to a lot
of copy-pasting. 

\section*{Discussion} 
Unfortunately we were unable to distribute the first set of questionnaires
before the project started, and another after the project. This meant that we
asked the students after completing the project to evaluate their knowledge
prior to starting the project. We believe that this could have an impact on the
responses.

One of the participating classes ended up not programming the \texttt{air:bit}
sensor kits themselves, but sharing a complete solution. We did not make it
possible to identify specific classes from individual responses, so we could not
exclude the responses from these students. These students will not have
completed all the different programming tasks, and we believe that they may
influence the results. 

\section*{Conclusions and Future Work} 
We have successfully deployed our \texttt{air:bit} project to schools across
Northern Norway. Doing so we have introduced more students to computer
programming and electronics. The new version of our sensor kit provides higher
quality datasets, and together with the online resources the kits are easier to
assemble and program. The results from the surveys show that students enjoy the
project and have positive learning outcomes. There are still students that
report that they are left with little knowledge on programming and fundamental
concepts in computer science after participating, and we aim to improve this.

One area of the project we have not discussed yet, is how to analyze the
collected air pollution measurements. This has been left out to the teachers,
but we are experimenting with interactive Jupyter\footnote{\url{jupyter.org}}
notebooks that allow students to interactively explore their data with
statistical programming languages such as R or Python. 

We are currently in the process of organizing a third round of the
\texttt{air:bit} project with even more participating schools. We aim to
continue improving the course contents and will use the responses from both
students and teachers to do so.

\section*{Acknowledgements} 
First we would like to thank all the students and teachers who participated in
the project.  We would like to thank \gls{met} for providing open weather
observations through \url{data.met.no} and \url{yr.no}. We would like to thank
\gls{nilu} for providing air quality measurements through their
\url{luftkvalitet.nilu.no} portal. We would also like to thank Thomas Olsen,
Sonja Grossberndt, and Juan Carlos Aviles Solis for their presentations and
input on the projects. We would like to thank the students at the Department of
Computer Science for their help teaching the course. Last, we would like to
thank Monica Martinussen and Ane Sætrum for their help developing the surveys. 

\bibliographystyle{ieeetr}
\bibliography{references}

\end{document}